# Commercialization of fuel cells: myth or reality?

Junye Wang
Faculty of Science and Technology
Athabasca University
1 University Drive, Athabasca, AB T9S 3A3, Canada
Telephone: 1-7803944883
Email: junyew@athabascau.ca

**Abstract**
　　Despite huge investment and efforts in the last decades, fuel cells are still known as a fledgling industry after 170 years of the first fuel cell. It becomes clear that these investment and efforts did not address the critical questions. Why upscaling of fuel cells failed often when many researchers stated their successes in small scale? Why the fuel cells with simpler structure still lag far from the internal combustion (IC) engines and gas turbines? Could the current investment of the hydrogen infrastructure reduce substantially the fuel cell cost and make a breakthrough to the key issues of durability, reliability and robustness? In this paper, we study these fundamental questions and point out a must-way possible to reduce cost of fuel cells and to substantially improve durability and reliability.

## 1. Introduction

　　Hydrogen can be renewable from water via electrolysis using electricity of solar, tidal or wind energy. Therefore, fuel cell and hydrogen technology have the potential in the comprehensive and balanced technology portfolio needed to address two most important energy challenges: significantly reducing carbon dioxide emissions and ending our dependence on fossil fuels. As a core technology of future hydrogen economy, fuel cells will play a pivotal role to revolutionize the way we power our world, offering cleaner, more-efficient alternatives to the internal-combustion (IC) engine in vehicles and the gas turbine at power station.
　　Although there are many successful special applications of fuel cells, such as unmanned aerial vehicles (UAV) and Apollo satellite, these special applications are not upmost target of fuel cell and hydrogen industries for the future hydrogen economy. Commercialization of a product is not if the product has been shipped successfully but if the product is profitable on the free and open market without any subsidies. Thus, if the cost and durability of fuel cells are not comparable to those of the IC engines or gas turbines, the fuel cells will not be accepted by market and the public. Clearly, over the last 50 years, a large number of institutes, private companies and government research labs have conducted serious R&D of fuel cell products. Progress has been incremental, leading to slightly better performance, reliability and durability. These may provide some benefit, but bring no end to the fuel cell itself. People are so wedded to these incremental improvements that may forget the upmost goals of R & D of fuel cells. In fact, the widespread adoption of fuel cells has yet to take off due to low durability and reliability and, in many cases, unacceptably high cost. So far, people have not seen any solution if fuel cells can replace IC engines according to both cost and durability in the near future. Some unrealistic promises by the research community who knew the goals of commercialization of fuel cells were unattainable created false expectations in the public community and investors who, once again, will see a deadline come and go without the huge victory promised. Fuel cell industry faces a precipice[1] in spite of many claimed successes.

## 2. A fledgling industry of 170 years history

The process of commercializing any new technology is fraught with a multitude of challenges. There are many highly successful products through improved technical developments and market deployment, such as airplanes, automobiles, computers, and cell phones. Therefore, in a fledgling industry, a large number of companies, investors and major governments have always believed that the technology of fuel cells was just a few years away from commercial success and a little more time and money would lead them to the ultimate breakthrough. However, the fuel cell technology does not follow other successful stories.

The history of fuel cells is comparable to that of the IC and steam engines. We have seen that the steam engines led industrial revolution and then the IC engines around the time of fuel cell exploration. An amount of US$22 billion has invested on R & D of fuel cell technology in Japan, USA and Europe over the past 18 years[2]. Payoffs from these efforts are reflected in the rapid growth in both the number of peer-reviewed publications and authorized patents. For instance, in Japan, the number of patents granted[2] to the companies in 2010 was 23 times as many as those in 2000. Today, fuel cells are still known as a fledgling industry after 170 years of the first invented fuel cell and massive investment. EU will launch new Framework Program, Horizon 2020 (2014-2020), in which €2.8 billion will be allocated to the fuel cell and hydrogen technology which leans heavily on the development of hydrogen infrastructure[3]. The ambitious goals are perfectly defensible, and indeed desirable, when we have the means to achieve them. However, the program is long on ambition and short on scientific detail. Unless one really understands those challenges in commercialization there is little chance of solving them. In fact, we have no any clear idea of both theoretical solution and technical measures how to solve durability, reliability and robustness in these programs. If public trust is not lost when advocates 'blueprints', it has been eroded when scientific infrastructure is unaccountable to the investors intended to benefit from its output; when there is not enough emphasis on translating research discoveries to the commercialization; and when target deadline cannot be achieved and marginal advances are over-hyped. Thus, we indeed need rethinking what are the real barriers of fuel cell upscaling.

## 3. The biggest barrier for commercialization of fuel cells

Generally speaking, the biggest barrier is still high cost (hydrogen and manufacturing) and technical issues (low robustness, reliability and durability) in commercialization of fuel cells since fuel cells are not comparable to the IC or turbine engines for both economic and technical conditions. The hydrogen price and infrastructure are considered to be a chicken-egg issue to solve high costs of fuel cells. Although commercialization of fuel cells will depend on R and D of whole hydrogen system as part of the future hydrogen economy, the importance of hydrogen price and infrastructure has been overestimated for market and public acceptance of fuel cell and hydrogen technology. Firstly, hydrogen is not unique fuel for fuel cells. There are many different types of fuel cells which do not use hydrogen as their fuels. If hydrogen is key problem, it is hard to explain the current dilemma of all other types of fuel cells. Secondly, hydrogen has no any relationship to manufacturing cost, robustness, durability and reliability. It is only related to run cost. Therefore, could the investment of the hydrogen infrastructure reduce significantly the fuel cell cost and make a breakthrough for the key issues of durability, reliability and robustness?

Due to a clear advantage of no moving-parts, fuel cells should be easier-fabrication and operation than the IC engines. Thus, the issue becomes another critical question why the fuel cells with simpler structures are more cost and lower reliability and durability? Although scientists and engineers did not answer the critical question, they still scaled them up for commercial applications from subsidies of governments. It was not surprising that they found that their fuel cells did not work well in the large scale due to technical issues of robustness, reliability and durability. Of course, the technical issues were still outward appearance because of degradation of materials or catalyst. No matter how many times they encountered problems of cost, durability, robustness or reliability. They always believed things could be fixed with other material, catalyst or sealing. Thus, R & D of fuel cell commercialization has been guided to solve issues of materials, chemistry, water and hotspot[4]. Overwhelming studies have been carried out for the fundamental issues of chemistry and materials as well as water and heat, and many measures have been suggested in industries, such as associated systems for water and heat management[5, 6], high temperature PEMFC[7], and cheaper catalyst[8, 9]. The cycling attempts of failure and improvement has been the most common approach which lead incremental progress, slightly better performance, reliability and durability. These may provide some benefit, but bring no end to the fuel cell itself. This is not to say that these researches are not important. They are important but overshadow the key barriers. Ironically, the efforts of the cycling attempts are not a joking but main practices in fuel cell areas for the past fifty years. Fuel cells need not tactical fixes but strategic solutions. Therefore, the real cause of upscaling failures has not been realized well yet in R & D of fuel cells.

It becomes clear that these past efforts do not address the critical questions. Why the simpler structure fuel cells still lag far from the steam and IC engines after massive investment in the past years? Why upscaling of fuel cells failed often when many researchers stated their successes in small scale? We have not seen any answer to the above basic questions from the past studies so far.

Fuel cells use matured electrochemical reactions of $H_2$ and $O_2$ which are much simpler than the combustion reaction of IC engine. Therefore, it is easy to make a single working fuel cell in the lab, but building fuel cell stacks that generate useful power reliably, efficiently, and cheaply is another matter entirely. The success of small scale cell means that the issues of chemistry, materials, water, and heat have been solved in single cell scale and has met technical requirements of durability, robustness and reliability. In order to obtain a higher voltage and current or power, many individual fuel cells are connected in either series or parallel, known as stacks. The upscaling technology using the repeat units are totally different from those of the conventional IC engines. This type of upscaling is based on a basic assumption of repeat units that performance of a successful cell can be repeated by all other cells since they use same materials, seals, catalyst and structures, and undertake same electrochemical processes.

In an assembly of repeat units, all the cells are designed to work in same operating conditions of fluid flow, electrochemical reaction and materials to maximize power output. However, there are main challenges to keep all the cells of the stack at same flow rate and pressure drop. Whilst some cells in the stack cannot reproduce performance of a designed successful cell, the outward appearance of the failure may be one of materials, chemistry and water and hotspot which lead a series of issues: durability, reliability and robustness. In practice, it is well-known that failure of a stack is usually because of the failure of some one individual cell, which leads the failure of the whole stack system. Obviously, the failed individual cell is generally because some cell works at non-designed operating conditions, such as higher temperature, even hotspot, due to local fast reaction. The hotspot temperature can exceed greatly the designed temperature of materials, leading to accelerated degradation or failure. The failure of catalyst may result from flooding and the failure of sealing may result from high mechanical stress due to high pressure difference. Thus, the failure of the upscaling is essentially because some cells deviate from design conditions in a stack. Such a deviation causes uneven electrochemical reaction. Particularly, the uneven reaction may be amplified due to blocking of some cells, leading serious degradation of materials and catalysts, water, heat or current issues. As a result, the system performance deteriorates totally the designed performance, reliability and durability.

Thus, all the failure of upscaling can be somehow attributed to the root cause of uneven flow distribution. Chemistry, materials, water and thermal issues are not essences but the outward appearance. As a result, the massive studies of materials, chemistry, water and thermal management have not been able to overcome issues of durability, reliability and robustness. It is poor flow-field designs that result in the failure of upscaling. The key to address the issue is to ensure all the cells work at its designed temperature condition rather than to develop a material at a higher temperature. It is questionable to address the upscaling failure using the cycling attempts of catalyst, water, heat, and chemistry and material issues. Surprisingly, these facts and figures are neither questioned nor discussed by expert groups and annual merit review[4, 10-11]. Also media representatives and other groups are quite disheartening in considering their immense significance in fuel cell community and investors[12].

## 4. Any solution?

The uniformity of flow distribution is a key for fuel cell upscaling. However, fuel cell companies and academics have ignored this key or very simply treated it. This may be because few industries have capacity to study systematically relationships of configurations, structure and function in flow field designs due to complexity of manifold systems or/and because academics take it as matured and solved technology. In fact, this has been a well-known challenge in flow distributions field for the fifty years[13-14]. It is like searching for a needle in a haystack to find the design which provides uniform flow distribution and proper pressure drop under a wide range of combinations in layouts, shapes, dimensions and operating conditions[15-17]. The time-effective design and optimization of flow fields have been one of the most challenges for cost reduction and performance improvement in fuel cells, which is recognized as a key breakthrough in R & D of fuel cells. Thus, the design of flow fields is the core technology in fuel cell upscaling.

Let's reach for what might in fact be possible. Ones might find an incredibly simple solution, which might make trials of commercialization shorter and less expensive. Encouragingly, Wang's research[15-18] addresses the key issue in upscaling fuel cells: 1) a completed theory of flow field designs and an effective tool of designs. Wang unified the main existing models and methods into one theoretical framework (e.g., main existing models in the theory as a special case), including Bernoulli and momentum theories, and discrete and continuum methods, 2) established a completed and comprehensive theory of flow field designs, 3) a direct, quantitative, and systematic relationship among flow distribution, pressure drop, configurations, structures, and flow conditions, and 4) developed an effective design tool with characteristic parameters, practical procedure, measures, criteria and guideline how to ensure all the cells/channels work in optimal operating windows.

Clearly, it is the must-way possible to substantially improve durability, reliability and robustness by using uniform flow distribution design with combining materials, catalyst, sealing and water and thermal management. Scientists and engineers need to work together to reorder priorities and change the conversation and culture of science and engineering. The process of increasing scientific understanding involves iteration between models, experiments, chemistry, materials, design and manufacturing. In this sense the theory of flow distribution stimulate our thinking and inform our experiments, including materials and chemistry, and point out the way toward a systematic assessment of flow distribution, materials, chemistry, configurations, structures and performances and their interactions. Thus, an interaction between research and development will result in greater success[16]. Dr Adamson[1] says that 'It really is do or die time in the industry. It is up to the industry itself to decide which scenario becomes reality.'